\documentclass{jpp}

\usepackage{graphicx,bm}
\usepackage{natbib}
\usepackage{color}
\newcommand{\bec}[1]{\mbox{\boldmath $ #1$}}
\newcommand{\meanUU}{\overline{\bm{U}}}
\newcommand{\meanWW}{\overline{\bm{W}}}
\newcommand{\meanBB}{\overline{\bm{B}}}
\newcommand{\meanU}{\overline{U}}
\newcommand{\meanW}{\overline{W}}
\newcommand{\meanS}{\overline{S}}

\title[Generation of large-scale vorticity in helical turbulence]{Generation of large-scale vorticity in rotating stratified turbulence with inhomogeneous helicity: mean-field theory}

\author[N. Kleeorin and I. Rogachevskii]%
{N.\ns K\ls L\ls E\ls E\ls O\ls R\ls I\ls N
  \and
I.\ns R\ls O\ls G\ls A\ls C\ls H\ls E\ls V\ls S\ls K\ls I\ls I
  \thanks{Email address for correspondence: gary@bgu.ac.il}
}
\affiliation{
Department of Mechanical Engineering, Ben-Gurion University of
the Negev, P. O. Box 653, 84105 Beer-Sheva, Israel
\\
Nordita, KTH Royal Institute of Technology
and Stockholm University, Roslagstullsbacken 23,
10691 Stockholm, Sweden
}

\date{\today; revised ; accepted  }
\begin{document}

\maketitle

%\preprint{NORDITA 2018-001}

\begin{abstract}
We discuss a mean-field theory of generation of large-scale vorticity in a rotating density stratified developed turbulence with inhomogeneous kinetic helicity.
We show that the large-scale nonuniform flow is produced due to ether a combined action of a density stratified rotating turbulence and uniform kinetic helicity or a combined effect
of a rotating incompressible turbulence and inhomogeneous kinetic helicity.
These effects result in the formation of a large-scale shear, and in turn its interaction with the small-scale turbulence causes an excitation of the large-scale instability (known as a vorticity dynamo) due to a combined effect of the large-scale shear and Reynolds
stress-induced generation of the mean vorticity.
The latter is due to the effect of large-scale shear on the Reynolds stress. A fast rotation suppresses this large-scale instability.
\end{abstract}

\section{Introduction}

A large-scale nonuniform flow or differential rotation
in a helical small-scale turbulence can result in generation of a large-scale
magnetic field by $\alpha\Omega$ or $\alpha^2\Omega$ mean-field dynamo
\citep[see, e.g.,][]{moffatt1978,parker1979,krause1980,zeldovich1983,ruzmaikin1988,ruediger2013},
where $\alpha$ is the kinetic $\alpha$ effect
and $\Omega$ is the angular velocity.
The kinetic $\alpha$ effect is related to a kinetic helicity
produced, e.g., by a combined action of uniform rotation and density stratified or
inhomogeneous turbulence.
Formation of the nonuniform flows is caused, e.g., by a rotating anisotropic
density stratified turbulence or turbulent convection.
The latter effect is also related to a problem of generation
of large-scale vorticity by a turbulent flow, and
has various applications in geophysical and astrophysical flows
\citep[see, e.g.,][]{lugt1983,pedlosky1987,chorin1994}.

It has been suggested by \cite{moiseev1983},
that the generation of the large-scale vorticity in a helical
turbulence occurs due to the kinetic alpha effect.
This idea is based on an analogy between the induction
equation for magnetic field and the vorticity equation \citep{batchelor1950}.
The latter implies that a large-scale instability
is associated with the term
${\bm \nabla} \times(\alpha \meanWW)$ in the equation for the mean vorticity,
$\meanWW$, similarly to the mean-field equation for the magnetic field, $\meanBB$,
where the key generation term is ${\bm \nabla} \times(\alpha \meanBB)$, see \cite{moiseev1983}.
A mean-field equation for the vorticity has been derived by \cite{khomenko1991}
using the functional technique for a compressible helical turbulence.
It has been shown there that the mean vorticity grows exponentially in time
due to the kinetic alpha effect.

However, the analogy between the induction
equation and the vorticity equation is not complete,
because the vorticity $\meanWW = {\bm \nabla} \times \meanUU$,
where the velocity $\meanUU$ is determined by the nonlinear
Navier-Stokes equation.
In addition, symmetry properties
imply that the term ${\bm \nabla} \times(\alpha \meanWW)$
in the mean vorticity equation should originate from
the Reynolds stress proportional to the mean velocity.
From one hand, the Reynolds stress enters neither the Navier-Stokes
nor the vorticity equations without spatial derivatives.
From another hand,
the Reynolds stress is an important turbulent characteristics, e.g.,
the trace of the Reynolds stress determines
the turbulent kinetic energy density.
To preserve the Galilean invariance,
the trace of the Reynolds stress as well as the diagonal components
of the Reynolds stress should be proportional to the spatial derivatives of
the mean velocity, rather than to the mean velocity itself.

\cite{frisch1987,frisch1988} have investigated the effect of
a non-Galilean invariant forcing that causes a large-scale
instability resulting in formation of a nonuniform
flow at large scales (so called the anisotropic kinetic alpha effect
or the AKA effect).
A non-Galilean invariant forcing and generation of
large-scale vorticity have been also investigated by \cite{kitchatinov1994}.
There are various examples for turbulence driven by non-Galilean
invariant forcing, e.g.,
supernova-driven turbulence in galaxies \citep{korpi1999}
and the turbulent wakes driven by galaxies moving
through the galaxy cluster \citep{ruzmaikin1989}.
Also presence of boundaries can break the Galilean
invariance, see e.g., discussion in \cite{brandenburg2001},
and references therein.

In a homogeneous non-helical and incompressible turbulence
with an imposed mean velocity shear, the large-scale vorticity
can be generated due to an excitation of a large-scale instability,
referred as a vorticity dynamo and caused by a combined
effect of the large-scale shear motions and Reynolds
stress-induced generation of perturbations of mean vorticity.
This effect has been studied
theoretically by \cite{elperin2003,elperin2007}
and detected in direct numerical simulations by \cite{yousef2008,kapyla2009}.
To derive the mean-field equation for the vorticity, the spectral
$\tau$ approach which is valid for large Reynolds numbers
has been applied by \cite{elperin2003}.
The linear stage of the large-scale instability which is saturated by nonlinear effects
has been investigated by \cite{elperin2003}, but not a finite time growth
of large-scale vorticity as described  by \cite{chkhetiany1994}.
In particular, a first order smoothing (a quasi-linear approach) has been used
in the latter study to derive equation
for the mean vorticity in a compressible random flow with an imposed
large-scale shear.
The latter approach is valid only for small Reynolds numbers, and
this is the reason why the large-scale instability has not been
found by \cite{chkhetiany1994}.
Importance of the vorticity dynamo has been demonstrated by \cite{hughes2015},
where they suggested a mechanism for the generation of large-scale magnetic
fields based on the formation of large-scale vortices
in rotating turbulent convection.

Formation of large-scale non-uniform flow by inhomogeneous helicity in a rotating
incompressible turbulence has been studied theoretically \citep{yokoi1993} and
in direct numerical simulations \citep{yokoi2016}.
The theoretical study and numerical simulations show that a non-uniform
large-scale flow is produced in the direction of angular velocity.
Recent direct numerical simulations have demonstrated formation
of large-scale vortices in rapidly rotating turbulent convection
for both, compressible \citep{chan2007,kapyla2011,mantere2011} and Boussinesq fluids
\citep{guervilly2014,rubio2014,favier2014}.
These large-scale flows consist of depth-invariant,
concentrated cyclonic vortices, which form by the merger of
convective thermal plumes and eventually grow to the size of
the computational domain. Weaker anticyclonic circulations
form in their surroundings.

In the present study we develop a mean-field theory of the
generation of large-scale vorticity in a rotating density
stratified turbulence with inhomogeneous helicity and large Reynolds numbers.
To derive the mean-field equation for the vorticity, the spectral
$\tau$ approach was applied here for a large Reynolds number turbulence.
We have shown that a non-uniform large-scale flow is produced in a rotating fully developed turbulence due to either inhomogeneous kinetic helicity or a combined effect of a density stratified flow and uniform kinetic helicity.
An interaction of the turbulence with the formed large-scale shear causes an excitation of the large-scale instability resulting in the generating of the mean vorticity (vorticity dynamo).
On the other hand, a fast rotation suppresses this large-scale instability.
The present study of the dynamics of large-scale vorticity in a rotating density
stratified helical turbulence demonstrates
that the mean-field equation for the large-scale vorticity
does not contain the ${\bm \nabla} \times(\alpha \meanWW)$ term
as was previously suggested by \cite{moiseev1983}.

\section{Effect of rotation on the Reynolds stress}

To study an effect of rotation on the Reynolds stress
in a rotating, density stratified and inhomogeneous
turbulence, we apply a mean-field approach and
use the Reynolds averaging.
In the framework of this approach, the
velocity and pressure are separated into
the mean and fluctuating parts.

\subsection{Equation for velocity fluctuations}

To determine the Reynolds stress, we use equation for fluctuations of
velocity ${\bm u}$, which is obtained by subtracting
equation for the mean field from the
corresponding equation for the instantaneous
field:
\begin{eqnarray}
{\partial {\bm u} \over \partial t} &=& - (\meanUU \cdot
\bec{\nabla}) {\bm u} - ({\bm u} \cdot \bec{\nabla}) \meanUU -
{\bec{\nabla} p \over \rho_{0}}
+ 2 {\bm u} \times {\bm \Omega} + {\bm U}^{N} .
\label{B2}
\end{eqnarray}
Here $p$ are fluctuations of fluid pressure, $\meanUU$ is the
mean fluid velocity, and Eq.~(\ref{B2}) is written in the reference
frame rotating with the angular velocity ${\bm
\Omega}$.
The fluid velocity for a low-Mach-number fluid
flow satisfies the continuity equation written
in the anelastic approximation: ${\rm div} \,
(\rho_0 \, \meanUU) = 0$ and ${\rm div} \,
(\rho_0 \, {\bm u}) = 0$. The mean fluid density and pressure with the
subscript $ "0" $ correspond to the hydrostatic basic reference state, given by
the equation: $\bec{\nabla} P_{0} = \rho_{0} {\bm g}$.
The nonlinear term ${\bm U}^{N}$
which includes the molecular viscous force, $\rho_0 \, {\bm F}_{\nu}({\bm u}) $,
is given by
\begin{eqnarray*}
U^{N} &=& \langle ({\bm u} \cdot \bec{\nabla}) {\bm u}
\rangle - ({\bm u} \cdot \bec{\nabla}) {\bm u} + {\bm
F}_{\nu}({\bm u}) .
\end{eqnarray*}

The derivation of the equation for the Reynolds stress includes the
following steps: (i) use new variable for fluctuations of velocity
${\bm v} = \sqrt{\rho_0} \, {\bm u}$; (ii) derivation of the equation
for the second moment of the velocity fluctuations $\langle v_i \, v_j \rangle$
in the ${\bm k}$ space;
(iii) application of the spectral $\tau$ approach (see Sect.~\ref{tau})
and solution of the derived equation for $\langle v_i \, v_j \rangle$ in the
${\bm k}$ space; (iv) returning to the physical
space to obtain formula for the Reynolds
stress as the function of the rotation rate $\Omega$.
Here the angular brackets denote ensemble averaging.

\subsection{Equation for the Reynolds stress}

We apply a multi-scale approach \citep{roberts1975}, i. e.,
the instantaneous two-point second-order correlation
function is determined as
\begin{eqnarray*}
\langle v_i({\bm x},t) \, v_j ({\bm  y},t) \rangle
&=& \int \,d{\bm k}_1 \, d{\bm k}_2 \,\,
\langle v_i({\bm k}_1,t) v_j({\bm k}_2,t)\rangle \, \exp
\big[i({\bm  k}_1 {\bm \cdot} {\bm x}
+{\bm k}_2 {\bm \cdot}{\bm y}) \big]
\nonumber\\
&=& \int f_{ij}({\bm k},{\bm R},t)  \exp(i {\bm k}
{\bm \cdot} {\bm r}) \,d {\bm k} ,
\end{eqnarray*}
where we use large-scale variables: ${\bm R} = ({\bm x}
+ {\bm y}) / 2$ and ${\bm K} = {\bm k}_1 + {\bm k}_2$;
and small-scale  variables: ${\bm r} = {\bm x} - {\bm y}$,
and ${\bm k} = ({\bm k}_1 - {\bm k}_2) / 2$.
Using Eq.~(\ref{B1}) derived
in Appendix~\ref{appA}, we obtain equation for the correlation function,
$f_{ij}({\bm k},{\bm R},t) = \int \langle v_i({\bm k}_1,t) v_j({\bm k}_2,t) \rangle
\exp(i {\bm K} {\bm \cdot} {\bm R}) \,d {\bm K}$:
\begin{eqnarray}
{\partial f_{ij}({\bm k},{\bm R},t) \over \partial t} &=& (I_{ijmn}^U +
L_{ijmn}^{\Omega}) f_{mn} + \hat{\cal N} f_{ij} ,
\label{B3}
\end{eqnarray}
where ${\bm k}_1 = {\bm k} + {\bm K} / 2$ and ${\bm k}_2 = -{\bm k} + {\bm K} / 2$,
\begin{eqnarray*}
I_{ijmn}^U &=& \biggl[2 k_{iq} \delta_{mp} \delta_{jn} + 2 k_{jq} \delta_{im}
\delta_{pn} - \delta_{im} \delta_{jq} \delta_{np}
- \delta_{iq} \delta_{jn} \delta_{mp} + \delta_{im}
\delta_{jn} k_{q} {\partial \over \partial k_{p}} \biggr] \nabla_{p}
\meanU_{q}
\\
&& - \delta_{im} \delta_{jn} \, [{\rm div} \, \meanUU + \meanUU
{\bm \cdot} \bec{\nabla}],
\\
L_{ijmn}^{\Omega} &=& \int \left[D_{im}^{\Omega}({\bm k}_1) \, \delta_{jn} +
D_{jn}^{\Omega}({\bm k}_2) \, \delta_{im}\right] \, \exp(i {\bm K} {\bm \cdot} {\bm R}) \,d {\bm K}, \quad D_{ij}^{\Omega}({\bm k}) = 2 \varepsilon_{ijm} \Omega_n k_{mn} .
\end{eqnarray*}
Here $ \delta_{ij} $ is the Kronecker tensor, $k_{ij} = k_i  k_j /k^2$ and $\varepsilon_{ijk}$ is the Levi-Civita tensor.
The correlation function $f_{ij}$ is proportional to the fluid density
$\rho_0({\bm R})$ and $\hat{\cal N} f_{ij}$ are the third-order
moments appearing due to the nonlinear terms:
\begin{eqnarray*}
\hat{\cal N} f_{ij} = \int \left[\langle P_{im}({\bm k}_1)
v^{N}_{m}({\bm k}_1) v_j({\bm k}_2) \rangle
+ \langle v_i({\bm k}_1) P_{jm}({\bm k}_2) v^{N}_{m}({\bm k}_2)
\rangle \right] \, \exp(i {\bm K} {\bm \cdot} {\bm R}) \,d {\bm K}.
\end{eqnarray*}

\subsection{$\tau$ approach}
\label{tau}

Equation~(\ref{B3}) for the second-order moment $f_{ij}({\bm k})$ contains high-order
moments and a closure problem arises \citep[see, e.g.,][]{monin2013,mccomb1990}.
To simplify the notations, we do not
show the dependencies on ${\bm R}$ and $t$ in the 
functions $f_{ij}({\bm k},{\bm R},t)$ and $\hat{\cal N} f_{ij}({\bm k},{\bm R},t)$.
We apply the spectral $\tau$ approximation,
or the third-order closure procedure \citep[see, e.g.,][]{orszag1970,pouquet1976,kleeorin1990,rogachevskii2004}. The spectral $\tau$ approximation postulates that the deviations of the
third-order-moment terms, $\hat{\cal N}f_{ij}({\bm k})$, from the
contributions to these terms afforded by the background turbulence, $\hat{\cal N}f_{ij}^{(0)}({\bm k})$, are expressed
through the similar deviations of the second moments, $f_{ij}({\bm
k}) - f_{ij}^{(0)}({\bm k})$:
\begin{eqnarray}
\hat{\cal N}f_{ij}({\bm k}) - \hat{\cal N}f_{ij}^{(0)}({\bm k}) = -
{f_{ij}({\bm k}) - f_{ij}^{(0)}({\bm k}) \over \tau_r(k)},
\label{B6}
\end{eqnarray}
where $\tau_r (k) $ is the
characteristic relaxation time of the statistical
moments, which  can be identified with the
correlation time $\tau(k)$ of the turbulent
velocity field for large Reynolds numbers.
In Eq.~(\ref{B6}) the quantities with the
superscript $(0)$ correspond to the background
turbulence (i.e., a turbulence with $\nabla_{i} \meanUU = 0)$.
We apply the $\tau$-approximation~(\ref{B6}) only to study the
deviations from the background turbulence, which is
assumed to be known (see below).
Validation of the $\tau$ approximation for different
situations has been performed in various numerical
simulations and analytical studies
\citep{brandenburg2005,rogachevskii2007,rogachevskii2011,rogachevskii2012,brandenburg2012,kapyla2012}.

\section{Effects of rotation and kinetic helicity on the Reynolds stress}

In this section we consider a combined effect of rotation and kinetic helicity
on the Reynolds stress. To this end we consider a model for the background
rotating helical turbulence.

\subsection{Model for the background turbulence}
\label{model}

We use the following model of the background rotating, density stratified,
and inhomogeneous turbulence with inhomogeneous kinetic helicity:
\begin{eqnarray}
f_{ij}^{(0)} &=&  {E(k)\, [1 + 2 k \, \varepsilon_u \, \delta(\hat{\bm k}
\cdot \hat {\bm \Omega})]\over
8 \pi \, k^2 \, (1 + \varepsilon_u)}
\, \biggl\{ \biggl[ \delta_{ij} - k_{ij} + {i \over k^2} \, \left(\tilde\lambda_i \,
k_j - \tilde\lambda_j \, k_i\right) \biggr] \rho_0 \, \langle {\bm u}^2 \rangle^{(0)}
\nonumber\\
&& - {i \over k^2} \, \Big[\varepsilon_{ijp} \, k_p + \left(\varepsilon_{ipq} k_{jp}
+ \varepsilon_{jpq} k_{ip}\right) \tilde\lambda_q\Big]\rho_0 \, \chi^{(0)} \biggr\} ,
\label{B15}
\end{eqnarray}
where $\hat {\bm \Omega}={\bm \Omega}/\Omega$, $\delta(x)$ is the Dirac delta function,
$\chi^{(0)}=\langle{\bm u} \,{\bf\cdot}  \,(\bec{\nabla}
{\bf \times} \, {\bm u})\rangle^{(0)}$ is the kinetic helicity,
$\bec{\tilde\lambda} = \bec{\lambda} - \bec{\nabla} / 2$,
$\bec{\lambda} = -(\bec{\nabla} \rho_0) / \rho_0$,
$\tau(k) = 2 \tau_\Omega \bar \tau(k)$,
the turbulent correlation time $\tau_\Omega$
is given below.

We assume that the background turbulence is the Kolmogorov type turbulence with
constant fluxes of energy and kinetic helicity over the spectrum,
i.e., the kinetic energy spectrum
$E(k) = - d \bar \tau(k) / dk$, the function $\bar \tau(k) =
(k / k_{0})^{1-q}$ with $1 < q < 3$ being the
exponent of the kinetic energy spectrum $(q =
5/3$ for Kolmogorov spectrum), $k_{0} = 1 /
\ell_{0}$ and $\ell_{0} $ is the integral scale of
turbulent motions.

To derive Eq.~(\ref{B15}) we use the following conditions:
\\
(i) the anelastic approximation: ${\rm div} \,
(\rho_0 \, {\bm u}) = 0$, which implies that $(ik_i^{(1)} - \lambda_i) f_{ij}^{(0)}({\bm k},{\bm K}) = 0$ and $(ik_j^{(2)} - \lambda_j) f_{ij}^{(0)} ({\bm k},{\bm K})= 0$, where
${\bm k}_1 \equiv {\bm k}^{(1)} = {\bm k} + {\bm K}/2$ and ${\bm k}_2 \equiv {\bm k}^{(2)} = - {\bm k} + {\bm K}/2$;
\\
(ii) $\int f_{ii}^{(0)} ({\bm k},{\bm K}) \exp \left[i {\bm K} \cdot {\bm R}\right] \, d {\bm k} \, d {\bm K} = \rho_0 \, \langle {\bm u}^2 \rangle^{(0)}$;
\\
(iii) $i \varepsilon_{ipj} \int k_p^{(2)} f_{ij}^{(0)} ({\bm k},{\bm K}) \exp \left[i {\bm K} \cdot {\bm R}\right] \, d {\bm k} \, d {\bm K} = \rho_0 \, \chi^{(0)}$;
\\
(iv) $f_{ij}^{(0)} ({\bm k},{\bm K}) = f_{ji}^{*(0)} ({\bm k},{\bm K}) = f_{ji}^{(0)} (-{\bm k},{\bm K})$.

To introduce anisotropy of the background turbulence due to rotation, we consider an anisotropic turbulence as a combination of a three-dimensional isotropic turbulence and two-dimensional turbulence in the plane perpendicular to the rotational axis. The degree of anisotropy $\varepsilon_u$ is defined as  the ratio of turbulent kinetic energies of two-dimensional to three-dimensional motions.
In this model we neglect effects which are
quadratic in ${\bm \lambda}$, ${\bm \nabla}\chi^{(0)}$ and
${\bm \nabla} \langle {\bm u}^2 \rangle^{(0)}$.
Different contributions in Eq.~(\ref{B15}) have been discussed by \cite{batchelor1953,elperin1995,radler2003}.

The effect of rotation on the turbulent
correlation time is described just by an heuristic argument.
In particular, we assume that
$\tau_\Omega^{-2} = \tau_0^{-2} + \Omega^2 /  C_\Omega^{2}$,
that yields:
\begin{eqnarray}
\tau_\Omega = {\tau_0 \over \left[1 + \left(C_\Omega^{-1}  \, \Omega \, \tau_0\right)^2\right]^{1/2}} .
\label{B12}
\end{eqnarray}
For a fast rotation, $\Omega \, \tau_0 \gg 1$, the
parameter $\Omega  \, \tau_\Omega$
tends to a finite value, $C_\Omega\sim 1$,  where $\tau_0 = \ell_{0} /
u_{0}$ and $u_0$ is the characteristic turbulent velocity at the integral scale $\ell_{0}$.

\subsection{The Reynolds stress in a rotating and helical turbulence}
\label{shear}

In this section we determine the contribution to the Reynolds stress
$f_{ij}^{(\Omega,\chi)}$ caused by either rotation and stratification in helical turbulence or
rotation and inhomogeneous kinetic helicity.
For a slow rotation, $\Omega \, \tau_0 \ll 1$, the function $f_{ij}^{(\Omega,\chi)}$
is given by
\begin{eqnarray}
f_{ij}^{(\Omega,\chi)} = \int \tau \, \Big[\tilde L_{ijmn} f^{(0,\tilde\lambda)}_{mn} + (L_{ijmn}^{\nabla} + L_{ijmn}^{\lambda})  \, f^{(0,\chi)}_{mn} \Big] \,d{\bm k} ,
\label{B18}
\end{eqnarray}
where we use Eq.~(\ref{B7}) derived in Appendix~\ref{appA},
the tensors $f^{(0,\chi)}_{ij} \propto \varepsilon_{ijp} \, k_p \chi^{(0)}$
and $f^{(0,\tilde\lambda)}_{mn} \propto \left(\varepsilon_{ipq} k_{jp}
+ \varepsilon_{jpq} k_{ip}\right) \tilde\lambda_q \chi^{(0)}$
determine corresponding terms in the model~(\ref{B15}) of the background turbulence,
$\bec{\tilde\lambda} = \bec{\lambda} - \bec{\nabla} / 2$, and all other definitions are given in Appendix~\ref{appA}.
After integration in ${\bm k}$ space in Eq.~(\ref{B18}), we obtain
contribution to the Reynolds stress $f_{ij}^{(\Omega,\chi)}$
caused by either rotation and stratification in helical turbulence or
rotation and inhomogeneous kinetic helicity
for a slow rotation, $\Omega \, \tau_0 \ll 1$:
\begin{eqnarray}
f_{ij}^{(\Omega,\chi)} = {(q-1) \over 2 q} \rho_0 \tau_0 \ell_0^2
\left[\Omega_i \lambda_j + \Omega_j \lambda_i + {4 \over 15}
\left(\Omega_i \nabla_j + \Omega_j \nabla_i\right) \right] \chi^{(0)} .
\label{B10}
\end{eqnarray}

For a fast rotation, $\Omega \, \tau_0 \gg 1$, the contribution to the Reynolds stress $f_{ij}^{(\Omega,\chi)}$ caused by either rotation and stratification in helical turbulence or
rotation and inhomogeneous kinetic helicity is given by
\begin{eqnarray}
f_{ij}^{(\Omega,\chi)} = \int \tau \, \Big[L_{ijmn}^{\nabla} + L_{ijmn}^{\lambda}\Big]  \, f^{(0,\chi)}_{mn}  \,d{\bm k} .
\label{B24}
\end{eqnarray}
After integration in ${\bm k}$ space in Eq.~(\ref{B24}), we obtain
\begin{eqnarray}
f_{ij}^{(\Omega,\chi)} =  C_\Omega {(q-1) \over 4 q} \rho_0 \, \ell_0^2
\left\{\hat \Omega_i \lambda_j + \hat \Omega_j \lambda_i + \hat \Omega_i \hat \Omega_j
\left[\hat {\bm \Omega} \cdot \left({\bm \lambda} + {\bm \nabla}\right)\right] \right\} \chi^{(0)} ,
\label{B11}
\end{eqnarray}
where $\hat {\bm \Omega}={\bm \Omega}/\Omega$.
In the derivation of Eq.~(\ref{B11}), we take into account that
the turbulent time $\tau_\Omega$ for a fast rotation, $\Omega \, \tau_0 \gg 1$
is determined by Eq.~(\ref{B12}). To integrate over the angles in ${\bm k}$-space
for a fast rotation, we use the integrals~(\ref{BBBB14}) given in Appendix~\ref{appA}.

\subsection{Formation of the mean velocity shear}

Let us consider the case when the angular velocity, ${\bm \Omega}=(0, \Omega, 0)$, is perpendicular to the density stratification axes, ${\bm \lambda}=(\lambda, 0, 0)$.
For simplicity, also consider the case when the gradient of the kinetic helicity is
parallel to ${\bm \lambda}$, i.e., ${\bm \nabla} \chi^{(0)}=(\nabla \chi^{(0)}, 0, 0)$.
In this case, $f_{xy}^{(\Omega,\chi)}(x)$ is only one nonzero contribution to the Reynolds stress $f_{ij}^{(\Omega,\chi)}$ caused by either rotation and stratification in helical turbulence or rotation and inhomogeneous kinetic helicity for $\Omega \, \tau_0 \ll 1$:
\begin{eqnarray}
f_{xy}^{(\Omega,\chi)}(x)=f_{yx}^{(\Omega,\chi)}(x) = {(q-1) \over 2 q} \rho_0(x) \, (\Omega \tau_0) \ell_0^2 \left(\lambda \, \chi^{(0)} + {4 \over 15} \nabla \chi^{(0)} \right) .
\label{B16}
\end{eqnarray}
The last term in Eq.~(\ref{B16}) is in agreement with Eq.~(30) of \cite{yokoi2016}.

The steady-state solution of the momentum equation for the $y$-component of the mean velocity $\meanUU^{(S)}$ reads:
\begin{eqnarray}
\nabla_x \left[\rho_0(x) \, \nu_{_{T}} \left(\nabla_x \meanU_y^{(S)}\right) - f_{yx}^{(\Omega,\chi)}(x) \right] = 0 ,
\label{BB17}
\end{eqnarray}
where $\nu_{_{T}} = (q+3) u_0 \ell_0 / 30$ is the turbulent viscosity \citep{elperin2002} and
we take into account that the gradient of the mean pressure along ${\bm \Omega}$ vanishes.
Integrating Eq.~(\ref{BB17}) over $x$, we determine the formed large-scale shear:
\begin{eqnarray}
\meanS\equiv\nabla_x \meanU_y^{(S)} = {f_{yx}^{(\Omega,\chi)} \over \rho_0 \, \nu_{_{T}}} = {15 (q-1) \over 2 q (q+3)}  \Omega \tau_0^2  \left(\lambda \, \chi^{(0)} + {4 \over 15} \nabla \chi^{(0)} \right) .
\label{B17}
\end{eqnarray}
It follows from Eq.~(\ref{B17}) that the large-scale shear is produced in rotating turbulence due to either inhomogeneous kinetic helicity or a combined action of a density stratified flows and uniform kinetic helicity.

In the present study we assume that shear does not affect the background turbulence.
For large values of the shear rate, the background turbulence  and turbulent correlation time can be affected by the shear.
In this case the quenching of the correlation time can be increased by the shear, i.e.,
the shear can decrease the correlation time \citep{zhou2017}.
The inclusion of these effects in the background turbulence is a subject of a separate study.
On the other hand, the solution of Eq.~(\ref{B3}) determines the deviations from the background turbulence, and the obtained solution of this equation yields Eq.~(\ref{B7}), that
describes the effect of shear on turbulence.

\section{Generation of the large-scale vorticity}

The formed large-scale shear $\meanS$ in a turbulent flow causes an excitation
of the large-scale instability resulting in the generation of the mean vorticity
due to the vorticity dynamo.
The linearized equation for the small perturbations of the
mean vorticity is given by
\begin{eqnarray}
{\partial \meanWW \over \partial t} = \bec{\nabla} {\bf \times} \left[\meanUU^{(S)} {\bf \times} \meanWW + \meanUU {\bf \times} \meanWW^{(S)}
+ 2 \meanUU {\bf \times} {\bm \Omega} + \rho_0^{-1} \, \left(\bec{\cal F}^{(S)}
+ \bec{\cal F}^{(\nu_{_{T}})} \right) \right] ,
\label{B8}
\end{eqnarray}
where $\meanUU$ and $\meanWW$ are perturbations of the mean velocity and mean vorticity,
while $\meanUU^{(S)} =(0, \meanS x, 0)$ and $\meanWW^{(S)}=(0, 0, \meanS)$ are the equilibrium mean velocity and mean vorticity related to the formed large-scale shear $\meanS$,
given by Eq.~(\ref{B17}).
Here ${\cal F}^{(S)}_i =- \nabla_j \left(\rho_0 \, \langle u_i \, u_j\rangle^{(S)}\right)$
is the effective force caused the shear effect on the Reynolds stress,
$\bec{\cal F}^{(\nu_{_{T}})}$ determines the turbulent viscosity, and we neglect small kinematic viscosity.
Let us consider for simplicity small perturbations of the
mean vorticity, $\meanWW(t,z) = (\meanW_x, \meanW_y, 0)$,
so that Eq.~(\ref{B8}) reads:
\begin{eqnarray}
{\partial \meanW_x \over \partial t} &=& \meanS \, \, \meanW_y +
\nu_{_{T}} \meanW''_x  ,
\label{B20}\\
{\partial \meanW_y \over \partial t} &=& - \beta \, \meanS \, \ell_0^2
\, \meanW''_x - 2 \Omega \lambda \meanU_x + \nu_{_{T}} \meanW''_y  .
\label{B5}
\end{eqnarray}
(see Appendix~\ref{appB}),
where $\meanW'_i = \nabla_z \meanW_i$ and the coefficient $\beta$ has been
determined by \cite{elperin2003}:
$\beta = 4 (2 q^2 - 47 q + 108) / 315$.
For Kolmogorov energy spectrum $(q=5/3)$, the coefficient $\beta= 0.45$.
In Eqs.~(\ref{B20}) and~(\ref{B5}) we take into account the Coriolis force
and the density stratification.
In the presence of the density stratification due to the gravity field
that is directed perpendicular to the angular velocity,
we can neglect a weak centrifugal force.
In Eq.~(\ref{B20}) we take into account that the characteristic scale of the
mean vorticity variations is much larger than the maximum scale of
turbulent motions $\ell_0$.
Since $\meanW_y =\meanU'_x$, Eq.~(\ref{B5}) can be rewritten as
\begin{eqnarray}
{\partial \meanW'_y \over \partial t} &=& - \beta \, \meanS \, \ell_0^2
\, \meanW'''_x - 2 \Omega \lambda \meanW_y + \nu_{_{T}} \meanW'''_y   .
\label{B21}
\end{eqnarray}

We seek for a solution of Eqs.~(\ref{B20}) and~(\ref{B21}) in the
form $ \propto \exp[\gamma t + i(\omega + K_z z)]$,
where the growth rate of the large-scale instability
and the frequency of the generated waves
are given by
\begin{eqnarray}
\gamma &=& \left[\beta \, (\meanS\, \ell_0 \, K_z)^2 - \left({\Omega \lambda \over K_z}\right)^2\right]^{1/2} - \nu_{_{T}} \, K_z^2  ,
\label{B22}\\
\omega &=& {\Omega \lambda \over K_z} .
\label{B23}
\end{eqnarray}
Equation~(\ref{B22}) implies that rotation in a density stratified
turbulence decreases the growth rate of the large-scale instability.
Since we consider the case when the angular velocity is perpendicular
to the wave vector ${\bm K}$ of the mean vorticity perturbations,
large-scale inertial waves are absent in the system.
In the absence of rotation and density stratification,
the expression~(\ref{B22}) for the growth rate of the large-scale instability
coincides with that obtained by \cite{elperin2003}.
Equation~(\ref{B23}) describes three-dimensional slow Rossby waves in rotating density
stratified flows which are similar to those studied by \cite{elperin2017},
see Eq. (28). We remind that the system considered in this study is a three-dimensional one, where the angular velocity, ${\bm \Omega}=(0, \Omega, 0)$, stratification, ${\bm \lambda}=(\lambda, 0, 0)$, and the wave number, ${\bm K}=(0,0, K_z)$, are perpendicular each other.

The mechanism of the large-scale instability studied here is as follows.
The first term, $\meanS \, \, \meanW_y$, in
Eq.~(\ref{B20}) describes a stretching of the mean vorticity component
$\meanW_y$ by non-uniform motions, which produces the component $\meanW_x$.
On the other hand, the first term, $- \beta \, \meanS \, l_0^2 \, \meanW''_x$,
in Eq.~(\ref{B5}) determines a Reynolds stress-induced
generation of perturbations of the mean vorticity $\meanW_y$
by turbulent Reynolds stresses. In particular, this
term is determined by $[\bec{\nabla} {\bf \times} (\rho_0^{-1} \, \bec{\cal F}^{(S)})]_y$,
where ${\cal F}^{(S)}_i$ describes the effective force caused the shear
effect on the Reynolds stress.
The growth rate of the instability is caused by a combined effect
of the sheared motions and the Reynolds stress-induced
generation of perturbations of the mean vorticity
\citep{elperin2003,elperin2007}.
On the other hand, the equilibrium large-scale shear is produced
either rotating turbulence and inhomogeneous kinetic helicity
or a combined effect of a density stratified rotating
turbulence and uniform kinetic helicity (see Sect.~\ref{shear}).

The physical explanation for why the rotation quenches the vorticity growth
is the following.
In the presence of the density stratified rotating turbulence,
there are three effects:
(i) the three-dimensional slow Rossby waves;
(ii) the Reynolds stress-induced generation of perturbations
of the mean vorticity $\meanW_y$;
(iii) turbulent viscosity which decreases both,
the energy of the Rossby waves and
the Reynolds stress-induced generation of perturbations
of the mean vorticity $\meanW_y$.
When rotation is fast, the Reynolds stress-induced generation of perturbations
of the mean vorticity $\meanW_y$ is suppressed.
A slow rotation just decreases the latter effect,
so there is a competition between the
generation of perturbations of the mean vorticity
and the Rossby waves.

Note that additional terms in Eqs.~(\ref{B20}) and~(\ref{B5})
caused by a combined effect of kinetic helicity and large-scale shear,
are much smaller than the terms which are taken into account in these equations.
The combined effects of the uniform kinetic helicity, rotation
and stratification or non-uniform kinetic helicity and rotation
are only important for the production of the background large-scale
velocity shear.

\section{Conclusions}

In the present study, the following effects are investigated:
(i) the effect of density stratification on the production of the large-scale vorticity
by the helical rotating turbulence;
(ii) the large-scale instability (vorticity dynamo)
suggested by \cite{elperin2003} for incompressible non-helical
turbulence with a large-scale shear, has been generalised for the
case of density stratified rotating and helical turbulence.
In particular, we show that
the large-scale flow is produced in rotating turbulence due to
inhomogeneous kinetic helicity or a combined action of a density stratified
flows and uniform kinetic helicity.
This results in the formation of a large-scale shear determined by
the balance between turbulent viscous force and the effective force caused by the
modification of Reynolds stress by either rotation and inhomogeneous kinetic helicity
or a combined action of rotation and a density stratified turbulence with
a uniform kinetic helicity.
This large-scale shear interacting with a turbulent flow results in an excitation of the large-scale instability generating the mean vorticity due to the vorticity dynamo,
while fast rotation suppresses this instability.

\begin{acknowledgements}
This work was supported in part by the Research Council of Norway
under the FRINATEK (grant No. 231444),
%IR added:
the Israel Science Foundation governed by the Israeli
Academy of Sciences (grant No. 1210/15),
%IR.
and the National Science Foundation under grant No. NSF PHY-1748958.
The authors acknowledge the hospitality of NORDITA
and the Kavli Institute for Theoretical Physics in Santa Barbara.
\end{acknowledgements}

\appendix
\section{Derivation and solution of the Reynolds-stress equation}
\label{appA}

In this Appendix we derive and solve the equation for the Reynolds-stress.
To this end, Eq.~(\ref{B2}) is rewritten in the new variable for
fluctuations of velocity ${\bm v}= \sqrt{\rho_0}\, {\bm u}$:
\begin{eqnarray}
{1 \over \sqrt{\rho_0}} {\partial {\bm v}({\bm x},t) \over
\partial t} &=& - \bec{\nabla} \biggl({p \over \rho_0}\biggr)
+ {1 \over \sqrt{\rho_0}} \biggl[2 {\bm v} {\bm \times} {\bm \Omega} -
({\bm v} \cdot \bec{\nabla}) \meanUU
- G^U \, {\bm v}\bigg] + {\bm v}^N,
\label{B4}
\end{eqnarray}
where $G^U = (1 / 2) \, {\rm div} \, \meanUU + \meanUU {\bm \cdot}
\bec{\nabla}$ and ${\bm v}^{N} $ are the nonlinear
terms which include the molecular viscous terms. The fluid velocity
fluctuations ${\bm v}$ satisfy the equation $\bec{\nabla} \cdot {\bm
v} = {\bm \lambda} \cdot {\bm v} / 2$, where ${\bm \lambda}
= -(\bec{\nabla} \rho_0) / \rho_0$.
To derive equation for the Reynolds-stress, we rewrite the momentum equation in
a Fourier space:
\begin{eqnarray}
{dv_i({\bm k}) \over dt} &=& [D_{im}^{\Omega}({\bm k}) +
J_{im}^U({\bm k})] v_m({\bm k}) + v_i^N({\bm k}) ,
\label{B1}
\end{eqnarray}
where
\begin{eqnarray*}
J_{ij}^U({\bm k}) &=& 2 k_{in} \nabla_{j} \meanU_{n} - \nabla_{j} \meanU_{i} -
\left[{1 \over 2} \, {\rm div} \, \meanUU + i (\meanUU
{\bm \cdot} {\bm k}) \right]  \, \delta_{ij} .
\end{eqnarray*}
To derive Eq.~(\ref{B1}), we multiply the momentum equation written
in ${\bm k}$-space by $P_{ij}({\bm k}) = \delta_{ij} - k_{ij} $
to exclude the pressure term from the equation of motion.
Here we also use the following identities:
\begin{eqnarray*}
&& \sqrt{\rho_0} \, \big[\bec{\nabla} {\bm \times} [\bec{\nabla}
{\bm \times} ({\bm u} {\bm \times} {\bm \Omega})] \big]
= \big({\bm \Omega}  {\bm \times} \bec{\nabla}^{(\lambda)}\big)
\, (\bec{\lambda} {\bm \cdot} {\bm v})
 + \big({\bm \Omega} {\bm \cdot} \bec{\nabla}^{(\lambda)}\big) \,
\big(\bec{\nabla}^{(\lambda)}  {\bm \times} {\bm v}\big),
\\
&& \sqrt{\rho_0} \, \big[\bec{\nabla} {\bm
\times} [\bec{\nabla}  {\bm \times} {\bm u}]
\big]_{\bm k} = - \big[\Lambda^2 \, \delta_{ij} -
\Lambda_i \lambda_j \big] v_j({\bm k}),
\end{eqnarray*}
where $\bec{\nabla}^{(\lambda)} = \bec{\nabla}  +
\bec{\lambda} /2$ and $\bec{\Lambda} = i {\bm k} + \bec{\lambda} /2$.

To derive equation for the Reynolds stress, we
apply a standard multi-scale approach \citep{roberts1975}
and use Eq.~(\ref{B1}). The equation for
\begin{eqnarray}
f_{ij}({\bm k},{\bm R},t) &=&
 \int \langle v_i({\bm k}_1, t) \, u_j({\bm k}_2,t) \rangle
\exp(i {\bm K} {\bm \cdot} {\bm R}) \,d {\bm  K},
\label{D2}
\end{eqnarray}
is given by Eq.~(\ref{B3}), where
${\bm k}_1 = {\bm k} + {\bm  K} / 2$,
${\bm k}_2 = - {\bm k} + {\bm  K} / 2$,
and
\begin{eqnarray*}
I_{ijmn}^U =\int \left[J^U_{im}({\bm k}_1) \, \delta_{jn} + J^U_{jn}({\bm k}_2) \, \delta_{im}
\right] \, \exp(i {\bm K} {\bm \cdot} {\bm R}) \,d {\bm K} .
\end{eqnarray*}

To solve Eq.~(\ref{B3}) we extract in tensor $L_{ijmn}^{\Omega}$  the parts which
depend on large-scale spatial derivatives
or the density stratification effects, i.e.,
\begin{eqnarray}
L_{ijmn}^{\Omega} = \tilde L_{ijmn}+
L_{ijmn}^\nabla  + L_{ijmn}^\lambda  + O(\lambda^2, \nabla^2) ,
\label{B19}
\end{eqnarray}
where
\begin{eqnarray*}
\tilde L_{ijmn} &=& 2 \, \Omega_q \,
(\varepsilon_{imp}  \, \delta_{jn} +
\varepsilon_{jnp} \, \delta_{im}) \, k_{pq},
\quad
L_{ijmn}^\nabla = - 2\,\Omega_q \,
(\varepsilon_{imp}  \, \delta_{jn} -
\varepsilon_{jnp} \, \delta_{im}) \,
k_{pq}^\nabla ,
\\
L_{ijmn}^\lambda &=& - 2\,\Omega_q \,
\Big[(\varepsilon_{imp}  \, \delta_{jn} -
\varepsilon_{jnp} \, \delta_{im}) \,
k_{pq}^\lambda +{i \over k^2} (\varepsilon_{ilq} \,
\delta_{jn}   \,\lambda_m- \varepsilon_{jlq} \,
\delta_{im}\,\lambda_n) \, k_{l} \Big],
\\
k_{ij}^\nabla &=& {i \over 2 k^2} \, [k_i \nabla_j
+ k_j \nabla_i - 2 k_{ij} ({\bm k} {\bm \cdot} \bec{\nabla})],
\quad
k_{ij}^\lambda = {i \over 2 k^2} \, [k_i \lambda_j
+ k_j \lambda_i - 2 k_{ij} ({\bm k} {\bm \cdot} \bec{\lambda})] .
\end{eqnarray*}
Equation~(\ref{B3}) in a steady state and after applying the spectral
$\tau$ approximation~(\ref{B6}), reads
\begin{eqnarray}
f_{ij}({\bm k}) &=& L_{ijmn}^{-1}
\big[f^{(0)}_{mn} + \tau \, (I_{mnpq}^U + L_{mnpq}^{\nabla}
+ L_{mnpq}^{\lambda})  \, f_{pq} \big],
\label{B7}
\end{eqnarray}
where we neglected  terms $\sim O(\nabla^2, \lambda^2)$.
Here the operator $L_{ijmn}^{-1}({\bm \Omega})$ is the
inverse of $\delta_{im} \delta_{jn} - \tau \,
\tilde L_{ijmn}$, and it is given by
\begin{eqnarray}
L_{ijmn}^{-1}({\bm \Omega}) &=& {1 \over 2} [B_1
\, \delta_{im} \delta_{jn} + B_2 \, k_{ijmn} +
B_3 \, (\varepsilon_{imp} \delta_{jn}
+ \varepsilon_{jnp} \delta_{im}) \hat k_p +
B_4 \, (\delta_{im} k_{jn} + \delta_{jn} k_{im})
\nonumber\\
&& + B_5 \, \varepsilon_{ipm} \varepsilon_{jqn}
k_{pq} + B_6 \, (\varepsilon_{imp} k_{jpn} +
\varepsilon_{jnp} k_{ipm}) ] ,
\label{B14}
\end{eqnarray}
where $\hat k_i = k_i / k$, $B_1 = 1 + \phi(2
\psi)$, $B_2 = B_1 + 2 - 4 \phi(\psi)$,
$B_3 = 2 \psi \, \phi(2 \psi)$, $B_4 = 2
\phi(\psi) - B_1$, $B_5 = 2 - B_1$, $B_6
= 2 \psi \, [\phi(\psi) - \phi(2 \psi)]$, $\phi(\psi) = 1 / (1
+ \psi^2)$ and $\psi = 2 \tau(k) \, ({\bm k}
\cdot {\bm \Omega}) / k $.
Note that for a slow rotation, $L_{ijmn}^{-1}({\bm \Omega})
= \delta_{im} \delta_{jn} + \tau \tilde L_{ijmn}$.

To integrate in Eq.~(\ref{B24}) over the angles in ${\bm k}$-space
for a fast rotation, we use the following integrals:
\begin{eqnarray}
&& \int k_{ij}^{\perp} \,d\varphi = \pi \delta_{ij}^{(2)},
\quad \int k_{ijmn}^{\perp} \,d\varphi = {\pi \over 4} \Delta_{ijmn}^{(2)},
\label{BBBB14}
\end{eqnarray}
where $\delta_{ij}^{(2)} \equiv P_{ij}(\Omega) = \delta_{ij} - \Omega_i \Omega_j /\Omega^2$
and $\Delta_{ijmn}^{(2)} = \delta_{ij}^{(2)}\delta_{mn}^{(2)}
+ \delta_{im}^{(2)} \delta_{jn}^{(2)}+ \delta_{in}^{(2)} \delta_{jm}^{(2)}$.

\section{Effect of shear on Reynolds stress}
\label{appB}

There are two effects of shear on Reynolds stress.
First effect is related to the contribution due to the turbulent viscosity:
$\langle u_i \, u_j\rangle^{(\nu_{_{T}})} = - 2 \nu_{_{T}} \, (\partial \meanU)_{ij}$, and the second contribution determines the Reynolds stress-induced generation of perturbations of mean vorticity by the effect of large-scale shear on turbulence \citep{elperin2003}:
\begin{eqnarray}
\langle u_i \, u_j\rangle^{(S)} = - l_0^2 \, [4C_1 \, M_{ij}+ C_2 \, (N_{ij} + H_{ij})
+ C_3 \, G_{ij}] ,
\label{C1}
\end{eqnarray}
where $\nu_{_{T}}$ is the turbulent viscosity,
$(\partial \meanU)_{ij} = (\nabla_i \meanU_{j} + \nabla_j \meanU_{i}) / 2$,
\begin{eqnarray*}
M_{ij} &=& (\partial \meanU^{(S)})_{im} ({\partial \meanU})_{mj} + (\partial \meanU^{(S)})_{jm} ({\partial \meanU})_{mi} , \quad
G_{ij} = \meanW^{(S)}_i \meanW_j + \meanW^{(S)}_j\meanW_i   ,
\\
H_{ij} &=& \meanW^{(S)}_n [\varepsilon_{nim} (\partial \meanU)_{mj} + \varepsilon_{njm} (\partial \meanU)_{mi}]   , \; \;  N_{ij} = \meanW_n [\varepsilon_{nim} (\partial
\meanU^{(S)})_{mj} + \varepsilon_{njm} (\partial \meanU^{(S)})_{mi}] ,
\end{eqnarray*}
$\meanUU$ and $\meanWW$ are perturbations of the mean velocity and mean vorticity,
while $\meanUU^{(S)} =(0, \meanS x, 0)$ and $\meanWW^{(s)}=(0, 0, \meanS)$ are the equilibrium mean velocity and mean vorticity related to shear $\meanS$,
the coefficients, $C_1 = 8 (q^2 - 13 q + 40) / 315$, $C_2 = 2 (6 - 7 q) /45$, $C_3 = - 2 (q + 2) / 45$, depend on the exponent of the energy spectrum.
When small perturbations of the mean velocity, $\meanUU(t,z) = (\meanU_x, \meanU_y, 0)$
and the mean vorticity, $\meanWW(t,z) = (\meanW_x, \meanW_y, 0)$, depend only on $z$,
the effective force $\rho_0^{-1} \, {\cal F}^{(S)}_i =- \nabla_j \langle u_i \, u_j\rangle^{(S)}$
is given by
\begin{eqnarray}
\rho_0^{-1} \, {\cal F}^{(S)}_i = - \meanS \, \ell_0^2 (\beta \, \meanW'_x, \beta_0 \, \meanW'_y, 0) ,
\label{C2}
\end{eqnarray}
where $\beta=C_1 + C_2 -C_3$ and $\beta_0=C_2/2 - C_1 -C_3$. Here we used the following identities:
\begin{eqnarray*}
\nabla_j M_{ij} &=& - (\meanS /4) (\meanW'_x, - \meanW'_y, 0) , \quad \nabla_j N_{ij} = - (\meanS /2) (\meanW'_x, 0, 0) ,
\\
\nabla_j H_{ij} &=& -(\meanS /2) (\meanW'_x, \meanW'_y, 0) , \quad \nabla_j G_{ij} = \meanS (\meanW'_x, \meanW'_y, 0) .
\end{eqnarray*}
Equation~(\ref{C2}) yields the first term in the right hand side of Eq.(\ref{B5}), see
Eq.(\ref{B8}).

%\newpage

\bibliographystyle{jpp}

\bibliography{Vorticity-inhomogeneous-helicity}

\end{document}